# High fidelity fibre-based physiological sensing deep in tissue

Tushar R. Choudhary[1,2,†], Michael G. Tanner[2,3,†], Alicia Megia-Fernandez[4], Kerrianne Harrington[5], Harry A. Wood[5], Adam Marshall[2], Patricia Zhu[4], Sunay V. Chankeshwara[4], Debaditya Choudhury[3], Graham Monro[2], Muhammed Ucuncu[4], Fei Yu[5], Rory R. Duncan[1], Robert R. Thomson[2,3], Kevin Dhaliwal[2], Mark Bradley[2,4]

†These authors contributed equally to this work

[1]Institute of Biological Chemistry, Biophysics and Bioengineering, School of Engineering & Physical Sciences, Heriot-Watt University, Edinburgh, UK

[2]EPSRC IRC Hub, MRC Centre for Inflammation Research, Queen's Medical Research Centre, University of Edinburgh, Edinburgh, UK

[3]SUPA, Institute of Photonics and Quantum Sciences, Heriot-Watt University, Edinburgh, UK

[4]*EaStChem,* School of Chemistry, University of Edinburgh, Edinburgh, UK

[5]Centre for Photonics and Photonic Materials, Department of Physics, University of Bath, Bath, UK

## ABSTRACT

Physiological sensing deep in tissue, remains a clinical challenge. Here a flexible miniaturised sensing optrode providing a platform to perform minimally invasive *in vivo in situ* measurements is reported. Silica microspheres covalently coupled with a high density of ratiometrically configured fluorophores were deposited into etched pits on the distal end of a 150 μm diameter multicore optical fibre. With this platform, multiplexed photonic measurements of pH and oxygen concentration with high precision in the distal alveolar space of the lung is reported. We demonstrated the phenomenon that high-density deposition of carboxyfluorescein covalently coupled to silica microspheres shows an inverse shift in fluorescence in response to varying pH. This platform delivered fast and accurate measurements, near instantaneous response time, no photobleaching, immunity to power fluctuations and a flexible architecture for addition of multiple sensors.

## INTRODUCTION

Alterations in the physiological environment in tissues can drastically impact biological processes. Despite the presumption of tightly regulated levels of key physiological parameters such as [H$^+$] and oxygen, in many areas of the human body the environmental physiology is unknown due to the paucity of miniaturised clinically compatible technologies. Thus, the aim of this study was to develop a flexible microendoscopic optrode for the accurate, robust and multiplexed sensing of pH and oxygen, which could be passed into remote regions of the human body. We demonstrate in this study that the distal alveolar acinar gas exchanging units of the lung can be accessed, where pH and oxygen play a critical role in maintaining homeostasis and are potential biomarkers of pathological processes, although clearly the platform technology is widely applicable to other regions.

The platform consists of novel pH sensors (fluorescein based) and oxygen sensors (palladium porphyrin complex based)[1] covalently attached to silica microspheres (10 μm diameter) loaded into pits etched into the distal facet of a 19 core (10 μm core diameter) multicore fibre (total diameter of ~150 μm, see Figure 2). Fluorescence is excited through selective coupling of light to a single core at the proximal end of the fibre, from which the spectrum is also measured, enabling multiplexing of sensors across the multi-core fibre.

## RESULTS

### Deposition of pH and oxygen sensors on multicore sensing platform

Following the pioneering work of Walt[2], fluorescent reporters covalently coupled to 10 μm amino modified silica microspheres were used as an optically robust solution to enable *in vivo* multiplexed pH and oxygen sensing. Thus, a bespoke multicore optical fibre (150 μm in diameter) with 19 germanium-doped cores matched to the diameter of the silica microspheres (10 μm) was fabricated and selectively etched with hydrofluoric acid, with the dopants directing the generation of concave pits aligned to the cores (see Figure 1). Into these pits the 10 μm silica microspheres were firmly and irreversibly deposited. This enabled each core to act as an isolated independent measurement channel, enabling multiparametric sensing through the specific illumination of different cores.

An important property of the amino modified silica microspheres is the local concentration of $1 \times 10^9$ free amino groups to which fluorescent sensors can bind. If all the sites are loaded with fluorophores, then the spacing between adjacent fluorophores would be ~5 Å. At this loading density following illumination, fluorophores would become excited with subsequent emission resulting in energy transfer with proximal neighbours (e.g. FRET) and as shown below this phenomenon which has never been previously reported, dramatically improves the performance of sensors.

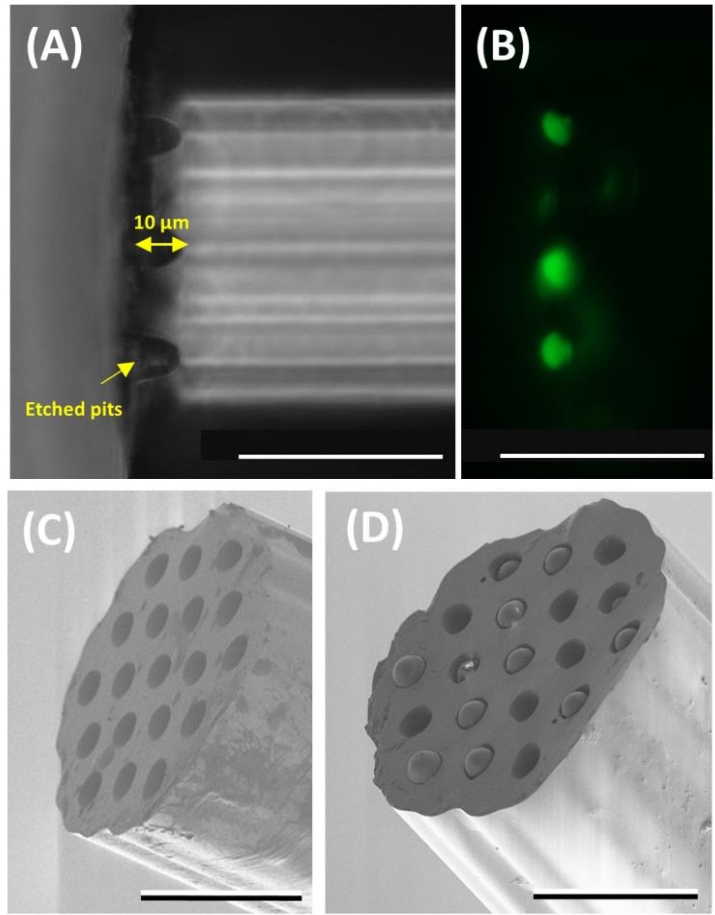

**Figure 1. Optrode fabrication and characterisation:** (A) Bright-field image of a fibre (viewed from the side) with etched pits (~10 μm in depth). (B) Fluorescence image (excitation 488 nm and emission 520 nm) of the etched fibres (viewed from the side) after the pH sensors (fluorescein-based) were loaded into the pits. Note the focal plane of image in (B) is different from (A) to highlight the loaded cores. (C) and (D) SEM images of an etched optical fibre before and after the addition of the microspheres. The scale bar in all the images is 50 μm.

Optrode measurements were performed using an epi-fluorescence arrangement (see Figure 2) with light selectively coupled into a single chosen core.

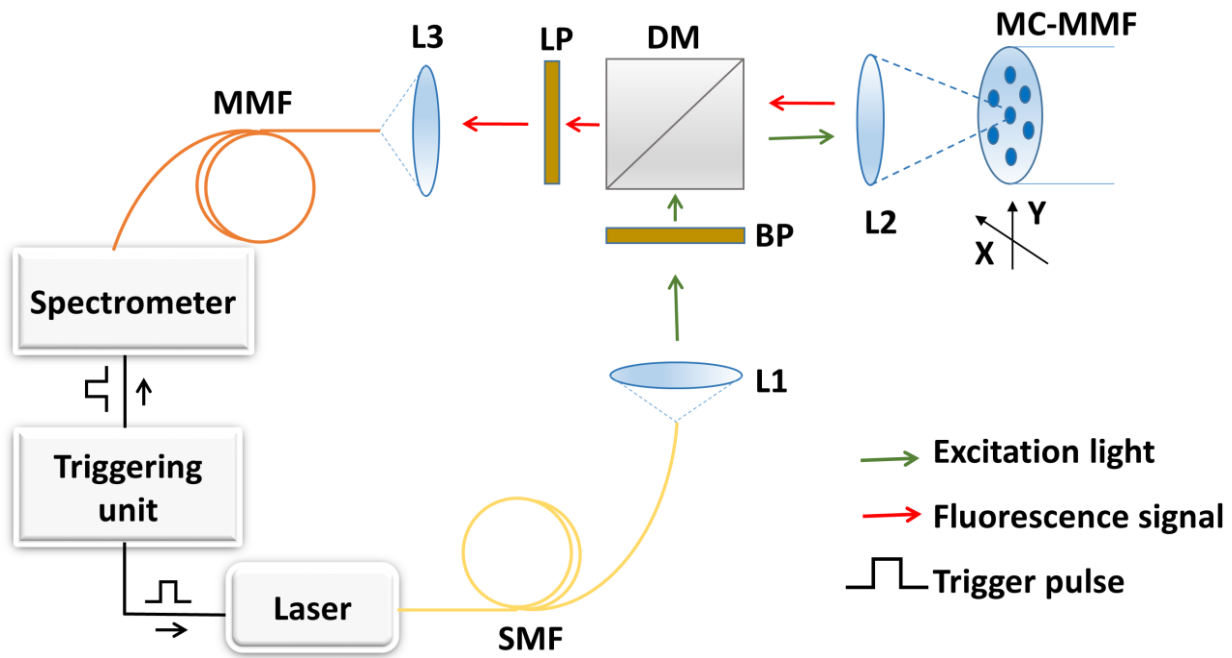

Figure 2. Optical setup to allow fibre based sensing: A 520 nm laser source (10 µW power and 100 ms exposure) was used as a pump source to excite both the pH and oxygen sensors. The pump illumination was launched from a single mode fibre (SMF) collimated by an aspheric lens (L1), passing through a band pass filter (BP) removing unwanted long wavelength light, and reflected by the dichroic mirror (DM), then focused with an identical aspheric lens (L2) and selectively coupled into the fibre core (with XY control of the sensing fibre mount). The returned light from the fibre core was collimated by L2, passed through the dichroic mirror and a long pass filter (LP, removing any reflected pump light). The fluorescence was then focused by lens L3 into a 50 µm core multimode fibre (MMF) patch cable, and directed to a spectrometer (Ocean Optics QEPro). The triggering unit (TTL pulse generator) controlled both laser and spectrometer allowing on demand short integration time (100ms) synchronised measurements.

*Silica microspheres loaded with fluorescent reporters for oxygen and pH sensing*

A widely used luminescent reporter of dissolved oxygen is the Palladium(II) 5,10,15,20-meso-tetrakis-(2,3,4,5,6-pentafluorophenyl)-porphyrin) complex (PdTFPP), that has good quantum yield and displays robust photostability[3,4]. PdTFPP absorbs between 400 nm and 560 nm, with fluorescent emission between 660 - 800 nm (see Figure 6)[5]. PdTFPP displays high fluorescence emission intensity in the region λ > 640 nm at low oxygen levels and the fluorescence levels reduce as oxygen levels increase. Importantly the fluorescent spectrum of PdTFPP has a non-variant region between 550 nm to 640 nm enabling ratiometric measurements (see Figure 6). An excitation wavelength of 520 nm was used, which was suitable for excitation of the pH sensors discussed below and also yielded reduced fibre auto fluorescent background signals.

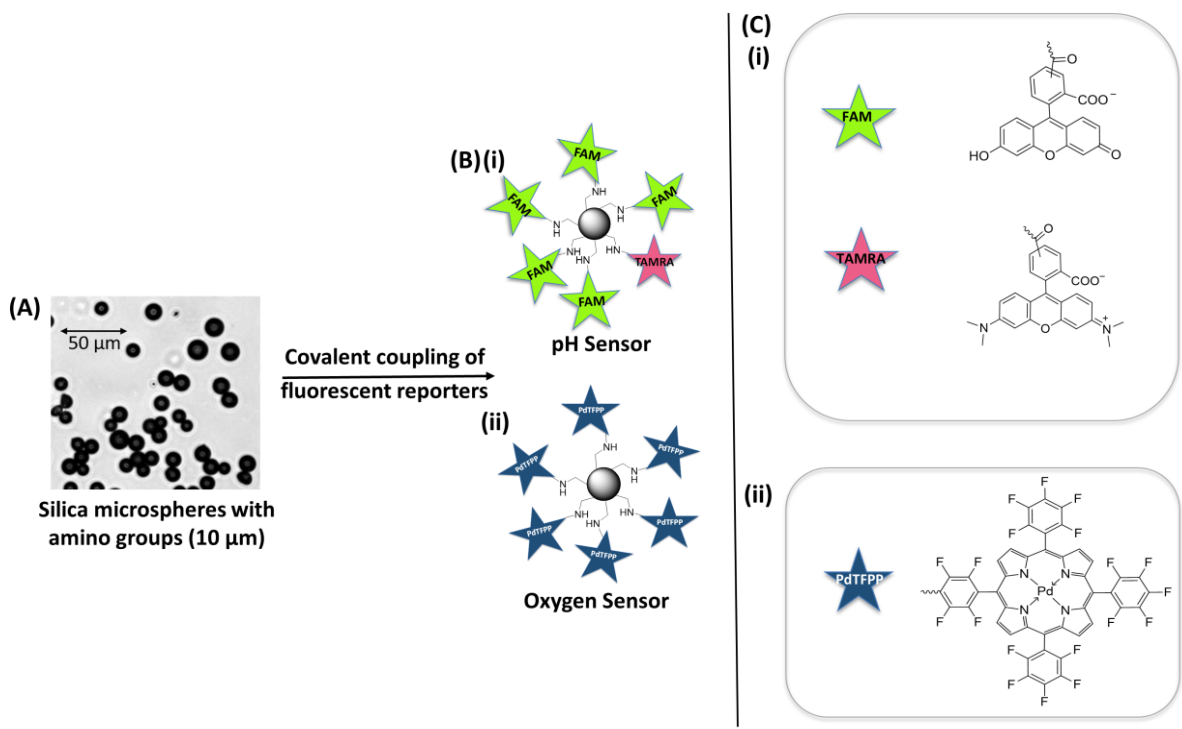

**Figure 3. Sensor fabrication: (A) Brightfield image of 10 μm amino functionalised silica microspheres. (B) (i) pH sensor: 5(6)-Carboxyfluorescein (FAM) and 5(6)-Carboxytetramethylrhodamine (TAMRA) covalently coupled onto the silica microspheres (not representative of loading ratio). (ii) Oxygen sensor: palladium(II) 5,10,15,20-meso-tetrakis-(2,3,4,5,6-pentafluorophenyl)-porphyrin (PdTFPP) covalently coupled onto the silica microspheres. (C) The structures of: (i) FAM and TAMRA and (ii) PdTFPP for conjugation onto the silica microspheres.**

Fluorescein is known to show increased emission intensity with increasing pH[6,7] which has led to fluorescein being a widely applied pH sensor in a variety of platforms.[8-12] The amino modified silica microspheres were saturation coupled with FAM (see Methods). However, contrary to all literature reports on fluorescein pH response, the microsphere sensors exhibited a totally unexpected decrease in emission with increasing pH (Figure 4 D), confirmed with conventional fluorescence microscopy of microspheres in pH buffers (Figure 4 B & C). Interestingly, when the loading was reduced (see the SI section for comparison) the sensors behaved as per literature expectations (SI Figure S1).

Sensors with different loading densities were observed to exhibit significantly different absorption spectra, with the microspheres displaying (macroscopically) very different properties (Figure 4 A). The "saturated" microspheres exhibited absorption features at shorter wavelength (Figure 4 E) suggesting, perhaps, the presence of dimer and/or trimers of fluorescein.[13] The "saturated" pH sensors showed lower emission intensity than the "standard" loaded sensors (see SI, Figure S1) due to an enhanced internal quenching effect since the dye molecules are in such close proximity on the microspheres. In addition, corresponding longer wavelength emission was observed for the "saturated" loaded pH sensors (see Figure S1). The "saturated" loaded pH sensors were also observed to be more robust to photo-bleaching (see SI, Figure S2), perhaps due to the promoted relaxation mechanisms between molecules in close proximity.

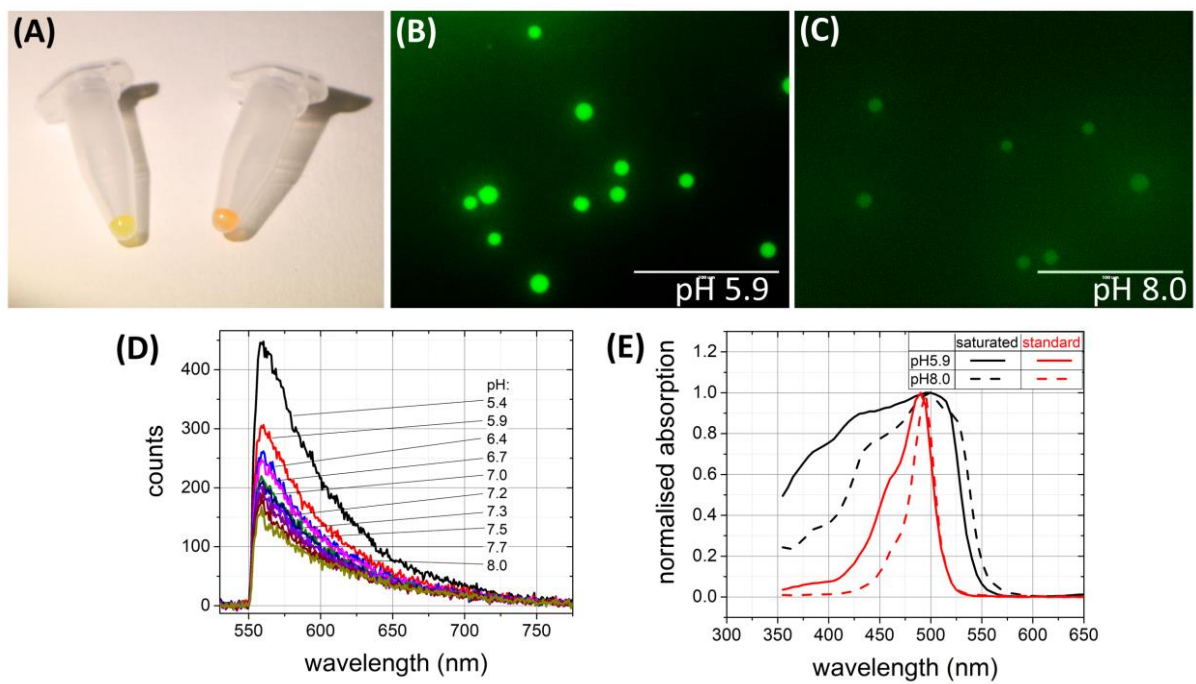

**Figure 4. Comparison of different pH sensors: (A) Image of "standard" loaded (left) and "saturated" loaded (right) pH sensors (note the colour difference). (B) and (C) "Saturated" pH sensors imaged via a fluorescence microscope at pH 5.9 (B) and pH 8.0 (C) (scale bar = 100 μm). (D) Emission spectra from the "saturated" pH sensors at different pH's. (E) Absorption spectra of the "saturated" and normally loaded pH sensors at pH 5.9 and 8.0.**

The fibre optrode utilising the "saturated" loaded FAM based pH sensor was observed to show a strong response to changes in pH and be much more robust than the "standard" loaded sensor. However, the intensity of a single fluorescent signal alone is always likely to be an unreliable measurement in a complex imaging or biological system, and thus a novel self-referencing pH sensor (details in next section) was designed.

**Ratiometric sensing**

Thus, 5(6)-Carboxytetramethylrhodamine (TAMRA) was attached along with 5(6)-Carboxyfluorescein (FAM), with the ratio optimised to allow observation of both spectra (see Figure 5). The spectral ratio to determine pH was calculated by dividing the number of counts in the TAMRA dominant region of the spectra (A2) by the number of counts in the FAM dominant region of the spectra (A1) (Figure 5). This fluorophore combination is a known FRET pair[14] and thus in combination with the "saturated" loading of both fluorophores leads to a complex relationship in the observed spectra, with the TAMRA fluorescent acceptor intensity changing in response to the fluorescein levels. The resulting dependence on pH of the combined spectral features differs in spectral regions dominated by the TAMRA (A2 region) or FAM (A1 region) in Figure 5 C. Figure 5 A shows the fluorescence intensity response of the "saturated" loaded FAM only pH sensor and TAMRA microsphere at different pH's. Only the FAM spectrum shows a significant change in intensity in response to pH.

In Figure 5 A & B it is apparent that the FAM spectra dominates region A1, while TAMRA dominates in region A2. When the fluorescence spectra of FAM and TAMRA were numerically combined (Figure 5 B, green line), the shape of the spectrum matched the FAM/TAMRA loaded probe (black line). The spectral signatures of the fluorophores were not fully resolvable from each other; however, a dominant peak was observed in the A2 region, originating from the TAMRA, with a "shoulder" in the A1 region from the FAM.

*pH measurements*
The FAM/TAMRA optrode response to pH is shown in Figure 5 C, D, with the analysed spectral ratio demonstrating responses that are robust, reversible and reliable to changes in pH with minimal variation between repetitions.

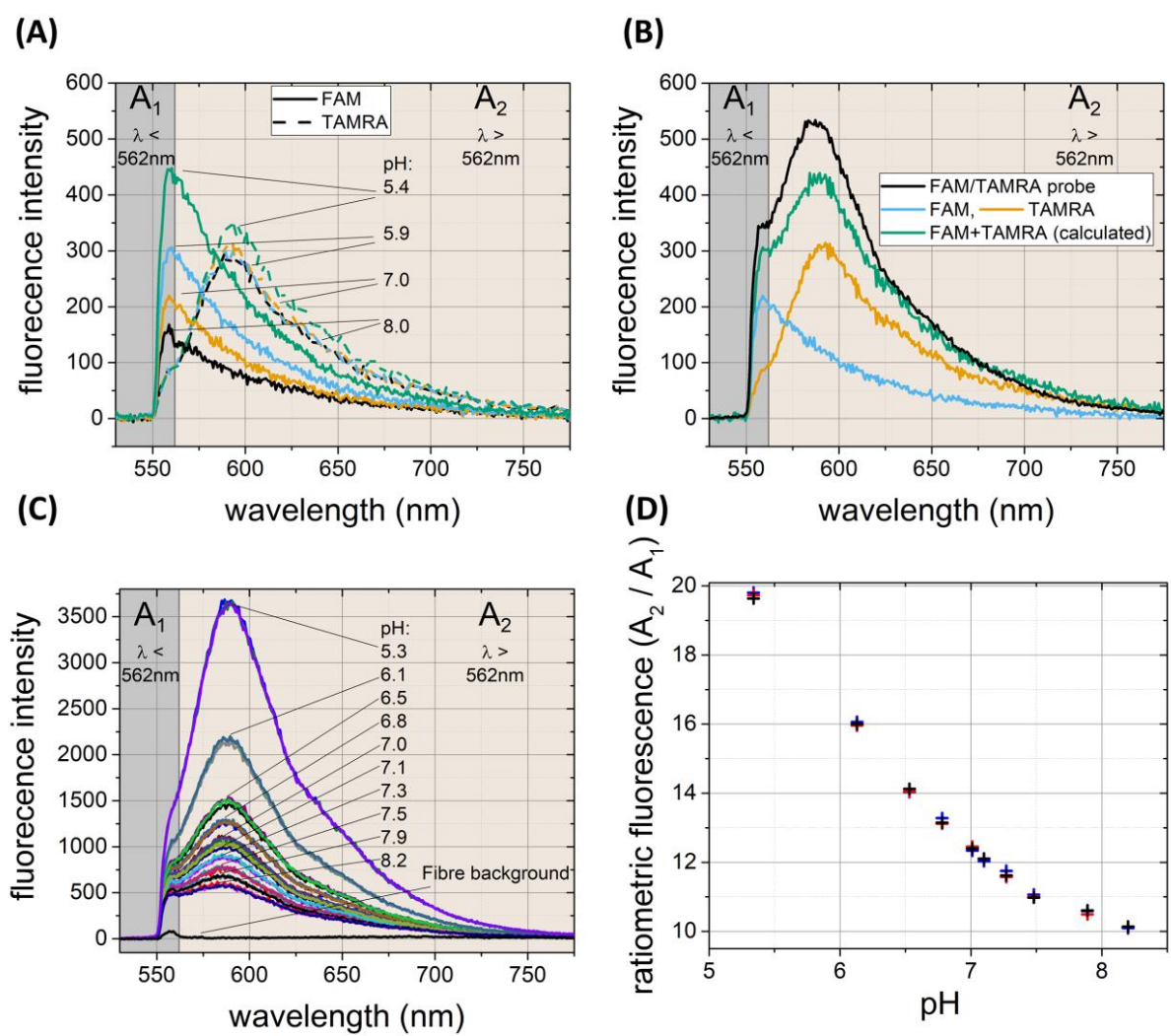

**Figure 5. pH response of the optrode and separate fluorophores. (A) Fluorescence emission spectra of "saturated" loaded FAM (solid lines) and TAMRA (dotted lines) microspheres at different pH's. (B) Fluorescence emission spectra of FAM only (blue line) and TAMRA microspheres (orange line) at pH 7. Fluorescence emission spectra at pH 7 of FAM and TAMRA microspheres when numerically combined (green line) is compared to the spectral**

shape of the fabricated FAM/TAMRA probe (black line, halved in amplitude to aid comparison). (C) Fluorescence emission spectra of the FAM/TAMRA pH sensor at different pH's. The FAM dominant region is highlighted as A1 ($\lambda < 562$ nm) and TAMRA dominant region as A2 ($\lambda > 562$ nm). (D) Calculated ratio of fluorescence intensity from the measured spectra at different pH's (n = 3). The ratio was calculated by dividing the area under the curve of spectral region A2 by A1. All measurements were from a single sensor from one core of the optrode at 520 nm (Power 10 µW and 100 ms exposure time).

*Oxygen concentration analysis*

The oxygen optrode (Palladium porphyrin complex) was used to measure dissolved oxygen concentrations with Figure 6 A showing the emission spectra of the oxygen sensor at various dissolved oxygen levels.

As previously noted, one region of the spectra was stable in fluorescence intensity with different oxygen concentrations, whilst another region changed with varying oxygen concentrations (highlighted as B1 and B2 in Figure 6 A). The ratiometric response (B2/B1) is shown in Figure 6 B.

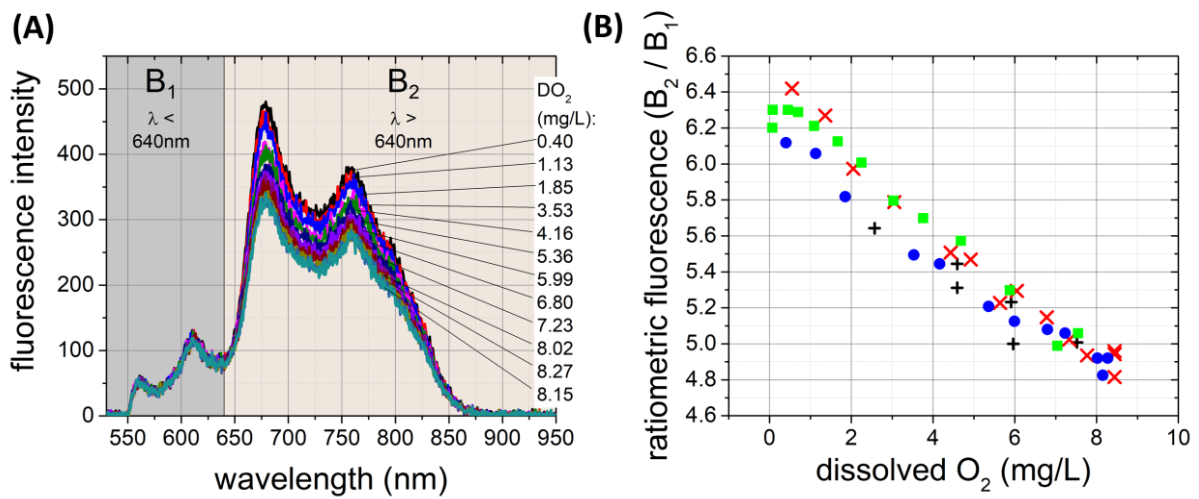

Figure 6. Response of the optrode to varying concentrations of dissolved oxygen. The optrode response was measured in distilled water changing the dissolved oxygen level from 8 mg/L (air saturated) to near full depletion (~0 mg/L). (A) Fluorescence emission spectra at different dissolved oxygen at 520 nm (Power 10 µW and 100 ms exposure time). The non-variant region is highlighted as B1 ($\lambda < 640$ nm) and variant region as B2 ($\lambda > 640$ nm). (B) Calculated ratio of fluorescence intensity from the measured spectra at different oxygen concentrations (n = 4). The ratio was calculated by dividing area under the curve in spectral region B2 by B1. All measurements were from a single sensor from one core of the optrode. Uncertainty in measurement was calculated to be ± 0.6 mg/L (standard deviation, see methods).

## Application in a perfused and ventilated *ex vivo* ovine lung model

The optrode was applied in a ventilated and perfused *ex vivo* ovine lung model,[15] with the fibre optrode passed trans-bronchially into the distal alveolar space. The perfusate was monitored for pH or $O_2$ using commercial meters while simultaneous measurements of these two analytes were measured with the optrode in the alveolar space throughout the duration of the experiment, with the pH or oxygen level of the circulating perfusate readily tuneable by the addition of base or altering the $O_2/N_2$ ratio in the ventilation circuit (see Figure 7). The optrode pH measurements correlated well with the commercial pH meter demonstrating its robustness and sensitivity in the whole lung model (see Figure 7 A). As expected, some difference was observed between alveolar tissue pH and perfusate pH, including a delay in tissue response to changing perfusate pH. The optrode $O_2$ measurements and the commercial meter also showed a good correlation to changes in the ventilated gas mixture (see Figure 7 B). As the lungs were undergoing ventilation with normal room air in addition to external gases, it is possible the fluctuations are representative of the true environment changing with each breath.

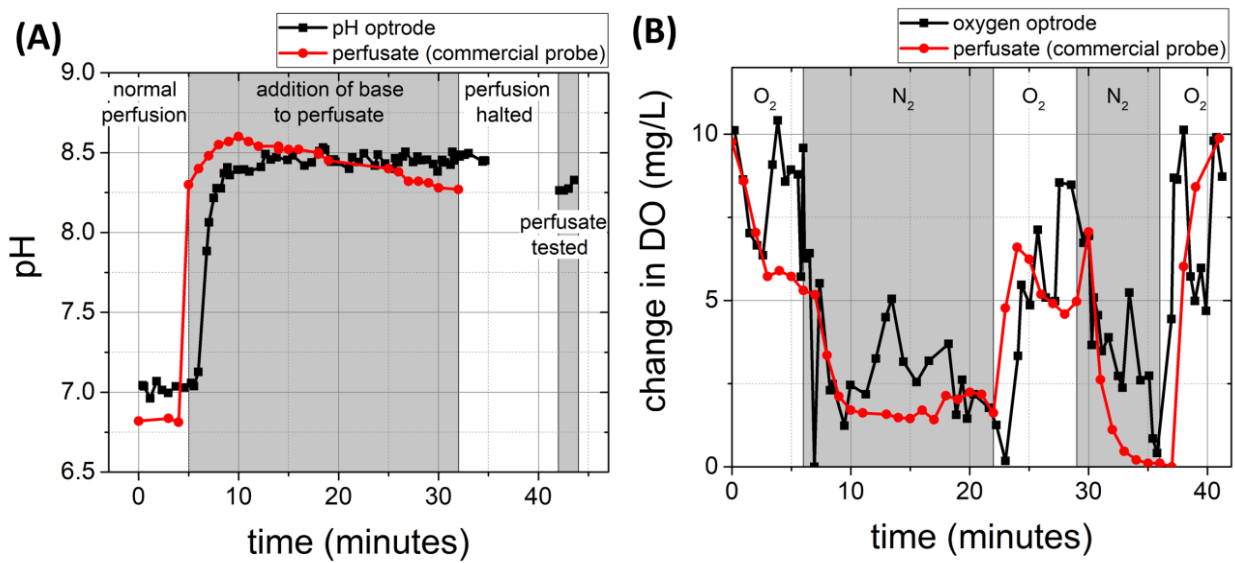

**Figure 7. Optrode measurements of pH and $O_2$ levels in a ventilated and perfused ovine lung. (A) The comparison of pH measurements taken with the optrode (black line) in the alveolar space and of the perfusate recorded with a conventional pH meter (red line) with pH modulation of the perfusate (addition of base). The data points on the right are from the optrode, which has been removed from the lung and placed in the perfusate. (B) Comparison of the change in dissolved oxygen measurement as determined with the optrode (black line) in the alveolar space and of the perfusate recorded with a commercial meter (red line), with switching of ventilated gas mixtures.**

## Optrode robustness

To demonstrate the photostability and reproducibility of the pH sensor, repeated measurements were taken using pH buffers (Figure 8 A). The ratiometric nature of the sensor means that it is resistant to variations in input laser power. To demonstrate this, the laser power was increased to a level where some photobleaching was observed (20 µW, double the optimal power, black crosses Figure 8 A), then reduced to 10 µW (optimised measurement power, red crosses Figure 8 A). Notably the ratiometric principle generated a measurement result unaffected by a 2-fold change in laser pump power, much greater than the fluctuations in the laser or optical coupling expected in normal operation. With normal operating conditions, photobleaching was minimal over the space of 25 measurements (Figure 8 A). See SI for investigation of the robustness of FAM "standard" and "saturated" loaded microspheres.

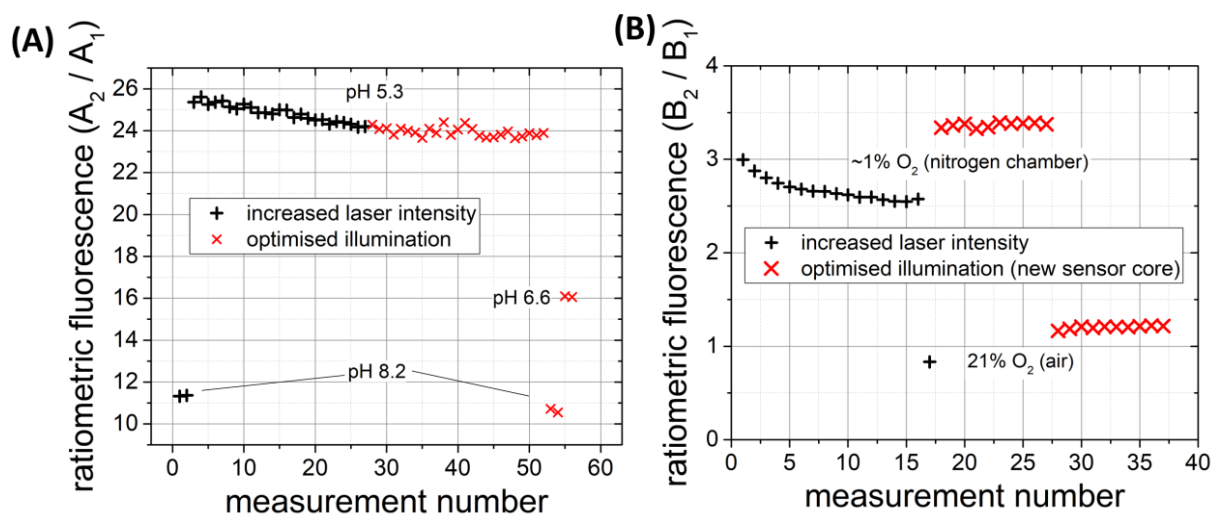

Figure 8. Measurement stability with repeated illumination and varying power: (A) Repeated measurement of pH with 20 µW illumination (black crosses), then reduced to 10 µW (red crosses). (B) Repeated measurement of oxygen concentration at 100 µW illumination (black crosses), then reduced to 10 µW (new sensor core, red crosses).

Similar tests were performed on the oxygen sensor to demonstrate photostability and immunity to changes in input laser power. To observe photobleaching, in this case, a greater increase in the input laser power of 10-fold was required (100 µW) compared to the optimised measurement power (10 µW). When a 10-fold higher illumination power was applied, some degradation was observed over 16 measurements shown in Figure 8 B. Reducing power to optimal levels confirmed no observable photobleaching effect on the measurement result over 20 repetitions (10 at depleted and 10 at room air oxygen levels).

## DISCUSSION

This work demonstrates high fidelity multiplexed measurement of pH and oxygen using a miniaturised flexible fibre based optrode. Exceptional measurement sensitivity was achieved for a miniaturised optrode, with an accuracy of the pH sensor of 0.02 pH units and oxygen sensor accuracy of 0.6 mg/L. Ratiometric pH measurements ensured robustness and resistance to the errors typical of intensity-only based measurements e.g. photobleaching, pump power fluctuations.  For the first time, we demonstrated that carboxyfluorescein covalently coupled to silica microspheres at high density exhibits an anomalous dependency of fluorescence intensity with pH, with increased pH reducing the fluorescence yield. Utilising triggered low power illumination and synchronised measurement, the fluorescent sensors are observed to be robust to photobleaching. The single miniaturised multicore fibre multiplexing approach demonstrated here offers the possibility of both pH and oxygen measurements *in vivo* at previously inaccessible sites.

Thus, we demonstrated the use of the optrode in a ventilated and perfused *ex vivo* ovine lung model, deployed into the air sacs, with both pH and oxygen sensors responding well to the change of pH and oxygen in the lung.

There are many reported optical fibre based sensors for either pH[8,9,16-20] or oxygen[21,22] which utilise trapping of nanoparticles/microspheres or sensor dye molecules on the fibre facet using suitable coatings such as sol-gel or polymers. The use of optical fibre in these techniques provides flexibility, but the sensors often have slow response times (in minutes) or limited accuracy and range. The use of a fluorescent polymer coating over an entire fibre facet is incompatible with precise deposition. Recently, two techniques for dual sensing of pH and oxygen have been described by the same group[23,24], in which a mixture of pH and oxygen sensing microsphere were deposited on the fibre using suitable sol-gel or hydrogel coating. The results demonstrated dual sensing, although the response time reported was slow (1.5-2 minutes for pH and ~30 seconds for oxygen) due to the coating deposition method. The multiplexed sensing platform described in this study overcomes these issues, and provides fast and accurate measurements, near instantaneous response time, no photobleaching, immunity to power fluctuations and a flexible architecture for addition of multiple sensors.

The use of multicore fibre also provides a highly flexible architecture. The work described here uses only two different sensors but the number of sensors can be increased easily with availability and requirement. The multiplexed fibre sensor assembly process is simple, low effort, robust, reliable, and the loaded microspheres self locate. The size of microspheres, size or number of fibre cores, or the size of the fibre itself can be very easily modified or adapted depending on the application requirement. The chemical loading of sensors (fluorophores) onto silica microspheres can simply be performed in a controlled, efficient and repeatable environment before being introduced to the optical fibres.

After successful demonstration of the multiplexed pH and oxygen sensor in *ex vivo* lung models, the next step will be to validate this technology through clinical translation, which will require packaging of the optrode in suitable biocompatible materials. Crucially, the optrode has been demonstrated to survive standard sterilisation procedures (see SI, Figure S3).

Our current understanding of key physiological parameters (pH, oxygen etc.) in health or in disease comes from invasive studies on animals[25] and cannot be applied to patients in clinic, either due to the invasive nature of the measurement or the lack of suitable techniques. The work described here is a significant step forward in the direction of clinical application of *in vivo* sensing in previously inaccessible areas of the human body which will help build our understanding of physiological parameters in disease pathology.

## METHODS

### Preparation of pH sensors

*FAM/TAMRA pH sensor*
Amino functionalized silica microspheres (10 µm, Kisker Biotech Gmbh & Co., 25 mg, 0.5 mL) were centrifuged, washed with MeOH and centrifuged (X3), washed with ether and centrifuged (x1) and left to dry at 40°C. A solution containing 5(6)-Carboxyfluorescein (FAM)/5(6)-Carboxytetramethylrhodamine (TAMRA) in a molar ratio of 300:1 (11.3 mg of 5(6)-Carboxyfluorescein and 0.04 mg of 5(6)-Carboxytetramethylrhodamine) and Oxyma (4.2 mg) in dimethylformamide (DMF, 0.42 mL) were stirred for 10 min, then N,N'-diisopropylcarbodiimide (DIC, 4.5 µL, 0.03 mmol) was added and after 1 minute the solution was added to the microspheres. The reaction was kept for 2 hours at 50°C. The reaction mixture was centrifuged and the supernatant carefully removed. The microspheres were washed with DMF and the coupling reaction was repeated three times. Finally, the microspheres were washed with DMF, 20 % Piperidine/DMF, DMF, MeOH, ether and dried at room temperature.

*"Saturated" and "Standard" loaded pH sensor*

The preparation method for the "saturated" loaded FAM only pH sensor was the same as above except that only FAM (11.4 mg) was used. In the case of the "standard" loaded pH sensor all the solutions were diluted by 10x.

*TAMRA loaded microspheres*

The same method as above for the FAM/TAMRA pH sensor was followed, except only TAMRA (13 mg) was used.

## Preparation of oxygen sensors

2 mg of PdTFPP (Sigma-Aldrich) and 50 mg of amino functionalised silica microspheres 10 µm (Kisker Biotech Gmbh & Co) were dispersed in 1 mL of diphenylether (Sigma-Aldrich). The reaction was heated under $N_2$ at 230°C in a vial for 3 hours. After cooling to room temperature, the mixture was centrifuged and the microspheres washed five times with dichloromethane to remove the diphenylether and the unreacted dye.

## Multicore multimode fibre fabrication

The multicore multimode fibre was fabricated using the "stack and draw" process commonly used to fabricate photonic crystal fibres as described in detail elsewhere[26]. In brief, a germanium doped optical fibre preform (diameter = 32 mm, peak numerical aperture = 0.3, Draka-Prysmian) with a parabolic refractive index profile and a thin pure silica jacket was drawn down to rods (ø = 5.75 mm). To increase core to core separation in the final fibre, each of the rods was further jacketed with a pure silica tube (outer diameter = 10 mm), and drawn down to a diameter of 2.4 mm. 19 rods of this material were then stacked in a hexagonal close-packed array, placed into a jacket tube, and drawn down under a vacuum to form the final 19 core fibre and the coating was added to the fibre. The final diameter of the cores was 10 µm, with a centre-to-centre separation of 23 µm. The outer diameter of the fibre was 150 µm.

## Optrode fabrication

The multicore fibre (19 cores, core diameter 10 µm) after removing the coating with a razor blade was cleaved to give the fibre flat ends. One end of the fibre was etched using 40 % hydrofluoric acid (HF) for 60 seconds. After etching, the end was sonicated in deionised water for 5 minutes to remove any HF from the cores. The etching parameters were optimised to produce pits with depths of ~10 µm to match the silica microspheres. The etched end of the fibre was then inserted into an eppendorf tube containing the silica microsphere loaded sensors (pH and oxygen) and gently tapped to load them onto the pits. The ends were then wiped using a tissue to remove any excess sensors.

## pH buffers

The pH buffers were prepared using a literature based protocol[15] with the buffer pH checked using a commercial pH meter (SevenGo Duo Pro, Mettler Toledo, U.K.).

## Dissolved oxygen variation

Oxygen concentration was measured in water saturated with 20 % oxygen (~8 mg/L) down to full depletion (0 mg/L). The concentration of oxygen in the distilled water was changed (between 0-8 mg/L) by bubbling either nitrogen or oxygen gas, and measured with a commercial $O_2$ meter (SevenGo Duo Pro, Mettler Toledo, U.K.) for every measurement taken with the optrode.

## *Ex Vivo* ovine lung model

Ovine lungs for the experiments described were from ewes destined for cull and were euthanized under Schedule 1 of Animals (Scientific Procedures) Act 1986, in an approved facility at the University of Edinburgh, performed in accordance with relevant guidelines and regulations. After barbiturate overdose and transection of major neck vessels, the trachea was dissected and clamped *in situ* and the heart and lungs were removed "en bloc". On the backtable, the heart was dissected off, leaving an open left atrium and a long section of pulmonary artery which was cannulated. The lungs were flushed with 2 litres of cold 0.9% saline with Heparin (2500 IU/L), retrograde flush via the pulmonary veins was then performed with a further 1 litre of cold saline. A 9-mm endotracheal tube (Rusch) was inserted into the trachea, secured and clamped with the lungs partially inflated. Lungs were then stored on ice for transportation and kept in a refrigerator (0-6˚C).

The *ex vivo* lung perfusion circuit consisted of an organ chamber to house the lungs connected to a hardshell reservoir (Maquet). The perfusate was circulated using a centrifugal pump (Maquet Rotaflow) back to the lungs via a gas membrane (Maquet) which was connected to a heat exchanger (Chalice Medical). The perfusate was delivered to the pulmonary circulation using a pulmonary artery catheter (XVIVO) with inline pressure sensing.

Once the lungs were transferred to the organ chamber 2 litres of room temperature perfusate (Phosphate Buffered Saline (PBSS, Gibco) supplemented with 10 % Foetal bovine serum (FBS, Gibco) and 10000 IU Heparin) were circulated. The pulmonary artery catheter was secured in the pulmonary artery using a silk tie and a digital temperature probe was sutured into a pulmonary vein. The lungs received a retrograde flush on the circuit to de-air the pulmonary artery cannula which was then connected to the circuit.

The heater was switched on and gentle flow was commenced at 0.25 L/min and increased gradually over 20 minutes to the desired flow rate (at least 50 % estimated cardiac output; 70 ml/kg/min) ensuring that the pulmonary artery pressure remained below 20 mmHg. Once the temperature of the effluent perfusate had reached 32°C, protective volume controlled ventilation was commenced at a tidal volume of 300 ml, positive end expiratory pressure 5, rate 8 beats per minute.

Bronchoscopy was performed once ventilation had commenced and a biocompatible Pebax tube (length 1.5 m) (Vention Medicals, U.S.) with an inner diameter of 1 mm and outer diameter of 1.4 mm was passed by transbronchial puncture to the alveolar space to a dependent segment of lung allowing deployment of the optrode.

An alkali solution (pH 8.5) was prepared using 40 % Trizma base (Sigma-Aldrich) in water for biasing of perfusate pH. $N_2$ and $O_2$ gases were introduced into the ventilation circuit (in addition to continued normal ventilation) to bias the respiratory oxygenation.

### pH measurements in lung model

The optrode was calibrated pre and post experiment in the pH buffers. The perfusate pH was tested directly with the optrode after removal from the lung, which showed good agreement with the commercial pH meter at the end of perfusion (Figure 7).

### Oxygen measurements in lung model

The optrode was calibrated in water (20 % oxygenated and depleted) pre and post experiment to correlate to a dissolved oxygen scale. However, a constant fluorescent background from the lung tissue contributed an offset in the calculated dissolved oxygen value thus data in Figure 8 for both perfusate measured with the commercial meter and from the optrode was background subtracted, such that change in dissolved oxygen (compared to the minimum observed) is plotted.

### Data processing and measurement uncertainty

Fluorescent spectra are analysed with simple code (Matlab). In the case of the $O_2$ spectra a fibre background was subtracted to better observe the signal and reference peak magnitude. Accuracy of measurement is calculated as follows. For pH measurement, as replicates were possible at repeatable pH values, standard deviation between the 3 replicates in Figure 5 D were calculated. In the region pH 6.5 to 7.5 the mean standard deviation was found. This was divided by the gradient of the response in this region (approximated as linear) to transpose to an error estimate on the x-axis, in pH units. For $O_2$ measurement, as it was not possible to precisely recreate a particular dissolved oxygen level, a slightly different approach was used. The ratio response in Figure 6 was approximated as linear, fitted, and the residuals found. The standard deviation of the residuals was calculated. This was transposed to the x-axis as for the pH measurement. As such both error calculations are quoted as the 1-sigma error.

# REFERENCES

1   Müller, B. J., Burger, T., Borisov, S. M. & Klimant, I. High performance optical trace oxygen sensors based on NIR-emitting benzoporphyrins covalently coupled to silicone matrixes. *Sensors and Actuators B: Chemical* **216**, 527-534, (2015).
2   Walt, D. R. & Michael, K. L.    (Google Patents, 2000).
3   Borisov, S. M., Lehner, P. & Klimant, I. Novel optical trace oxygen sensors based on platinum(II) and palladium(II) complexes with 5,10,15,20-meso-tetrakis-(2,3,4,5,6-pentafluorphenyl)-porphyrin covalently immobilized on silica-gel particles. *Analytica Chimica Acta* **690**, 108-115, (2011).
4   Koren, K., Borisov, S. M. & Klimant, I. Stable optical oxygen sensing materials based on click-coupling of fluorinated platinum(II) and palladium(II) porphyrins—A convenient way to eliminate dye migration and leaching. *Sensors and Actuators. B, Chemical* **169**, 173-181, (2012).

5       Quaranta, M., Borisov, S. M. & Klimant, I. Indicators for optical oxygen sensors. *Bioanalytical Reviews* **4**, 115-157, (2012).
6       Martin, M. M. & Lindqvist, L. The pH dependence of fluorescein fluorescence. *Journal of Luminescence* **10**, 381-390, (1975).
7       Mordon, S., Devoisselle, J. M. & Maunoury, V. in vivo pH measurement and imaging of tumor tissue using a pH-sensitive fluorescent probe (5,6–carboxyfluorescein): instrumental and experimental studies. *Photochemistry and photobiology* **60**, 274-279, (1994).
8       Dafu, C. *et al.* Optical-fibre pH sensor. *Sensors and Actuators B: Chemical* **12**, 29-32, (1993).
9       Islam, S., Bidin, N., Riaz, S., Krishnan, G. & Naseem, S. Sol–gel based fiber optic pH nanosensor: Structural and sensing properties. *Sensors and Actuators A: Physical* **238**, 8-18, (2016).
10      Stich, M. I. J., Fischer, L. H. & Wolfbeis, O. S. Multiple fluorescent chemical sensing and imaging. *Chemical Society Reviews* **39**, 3102-3114, (2010).
11      Bradley, M. *et al.* pH sensing in living cells using fluorescent microspheres. *Bioorganic & Medicinal Chemistry Letters* **18**, 313-317, (2008).
12      Han, J. & Burgess, K. Fluorescent Indicators for Intracellular pH. *Chemical Reviews* **110**, 2709-2728, (2010).
13      Rohatgi, K. K. & Singhal, G. S. Nature of Bonding in Dye Aggregates. *The Journal of Physical Chemistry* **70**, 1695-1701, (1966).
14      Caruso, F., Donath, E. & Möhwald, H. Influence of Polyelectrolyte Multilayer Coatings on Förster Resonance Energy Transfer between 6-Carboxyfluorescein and Rhodamine B-Labeled Particles in Aqueous Solution. *The Journal of Physical Chemistry B* **102**, 2011-2016, (1998).
15      Choudhury, D. *et al.* Endoscopic sensing of alveolar pH. *Biomedical optics express* **8**, 243-259, (2017).
16      Gu, B., Yin, M.-J., Zhang, A. P., Qian, J.-W. & He, S. Low-cost high-performance fiber-optic pH sensor based on thin-core fiber modal interferometer. *Opt. Express* **17**, 22296-22302, (2009).
17      Shao, L.-Y., Yin, M.-J., Tam, H.-Y. & Albert, J. Fiber optic pH sensor with self-assembled polymer multilayer nanocoatings. *Sensors* **13**, 1425-1434, (2013).
18      Lee, S. T. *et al.* A sensitive fibre optic pH sensor using multiple sol-gel coatings. *Journal of Optics A: Pure and Applied Optics* **3**, 355, (2001).
19      Mohamad, F. *et al.* Controlled core-to-core photo-polymerisation - fabrication of an optical fibre-based pH sensor. *Analyst* **142**, 3569-3572, (2017).
20      Ehrlich, K. *et al.* pH sensing through a single optical fibre using SERS and CMOS SPAD line arrays. *Opt. Express* **25**, 30976-30986, (2017).
21      MacCraith, B. D. *et al.* Fibre optic oxygen sensor based on fluorescence quenching of evanescent-wave excited ruthenium complexes in sol-gel derived porous coatings. *Analyst* **118**, 385-388, (1993).
22      Chu, C.-S. & Chuang, C.-Y. Optical fiber sensor for dual sensing of dissolved oxygen and Cu2+ ions based on PdTFPP/CdSe embedded in sol–gel matrix. *Sensors and Actuators B: Chemical* **209**, 94-99, (2015).
23      Vasylevska, G. S., Borisov, S. M., Krause, C. & Wolfbeis, O. S. Indicator-Loaded Permeation-Selective Microbeads for Use in Fiber Optic Simultaneous Sensing of pH and Dissolved Oxygen. *Chemistry of Materials* **18**, 4609-4616, (2006).
24      Kocincova, A. S., Borisov, S. M., Krause, C. & Wolfbeis, O. S. Fiber-Optic Microsensors for Simultaneous Sensing of Oxygen and pH, and of Oxygen and Temperature. *Analytical Chemistry* **79**, 8486-8493, (2007).
25      Shah, V. S. *et al.* Airway acidification initiates host defense abnormalities in cystic fibrosis mice. *Science* **351**, 503-507, (2016).
26      Russell, P. Photonic crystal fibers. *Science* **299**, 358-362, (2003).

## Acknowledgements

The authors acknowledge the UK Engineering and Physical Sciences Research Council (EP/K03197X/1) for funding this work.

The authors thank Gareth Williams for technical advice.

## Author contributions

TRC and MGT performed the instrument development, sensor characterisation, integrated the multiplexed fibre sensing architecture (optrode), performed the optrode characterisation and implementation, data analysis and prepared the manuscript. MB conceived the concept of the pH and oxygen sensors using microspheres for attachment on fibre. AMF and MU fabricated the pH sensors. TRC, MGT and AMF discovered and optimised the reverse (fluorescein effect) behaviour of the pH sensor. AM designed the ventilated and perfused lung model. AM, TRC, MGT and GM performed the lung experiments. KH, HW and FY designed and fabricated 19-core multimode fibre. SC and PZ fabricated the oxygen sensors. PZ performed oxygen sensor characterisation. DC and TRC optimised the etching parameters. DC etched the fibres. MGT, RRD, RRT, MB and KD provided strategic oversight and contributed to the manuscript. All authors discussed the results and commented on the manuscript.

# SUPPLEMENTARY INFORMATION

## Reverse pH sensor

Figure S1 shows effects of "saturated" and "standard" loading of FAM on pH sensors response. Figure S1 A shows the response of "saturated" loaded pH sensors at two different pH values compared to "standard" loaded shown in Figure S1 B. Corresponding longer wavelength emission is observed in a comparison of spectra with differing dye loading (Figure S1 C).

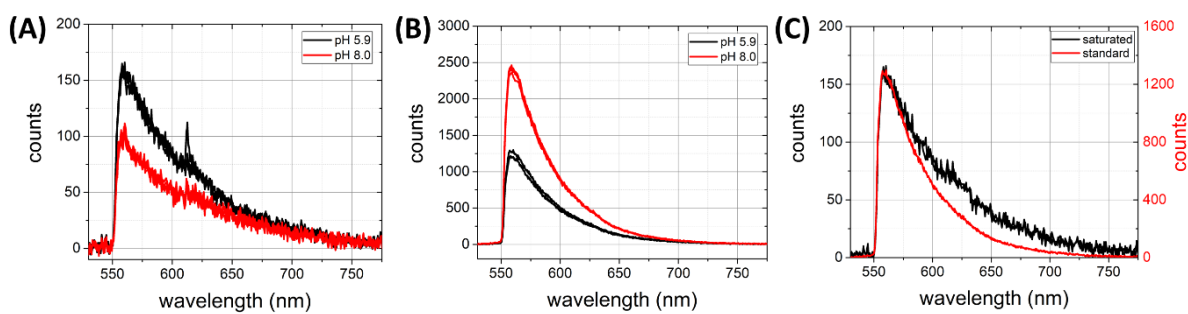

**Figure S1. (A)** FAM covalently attached to silica microsphere when "saturated" loaded decrease in fluorescence intensity with the increase in pH. **(B)** FAM covalently attached to silica microsphere when "standard" loaded increase in fluorescence intensity with increase in pH. **(C)** "saturated" (black line) and "standard" (red line) loading of FAM on silica microsphere have slightly different emission.

The "saturated" loading of FAM on silica microspheres was found to be advantageous as the resulting pH sensor was more robust to photobleaching (Figure S2 A) as compared to "standard" loading of FAM on silica microspheres (Figure S2 B)

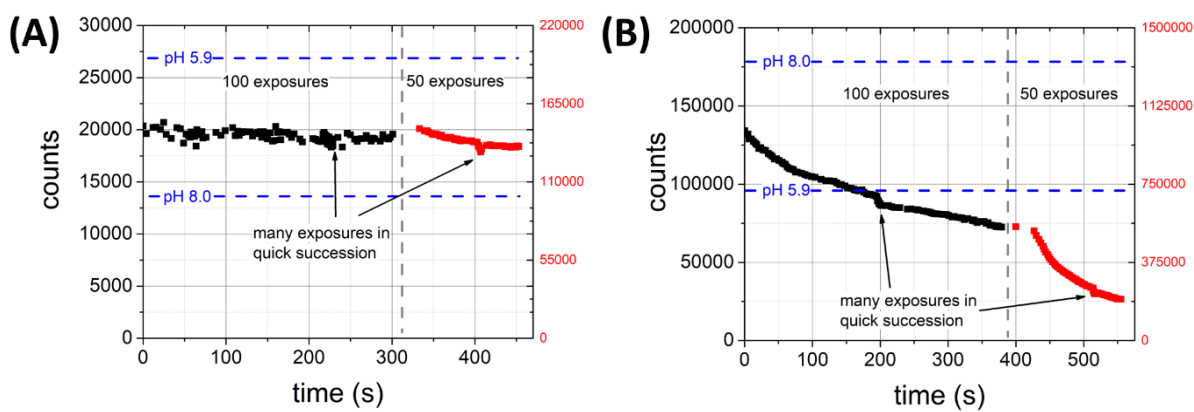

**Figure S2. (A)** "Saturated" loaded FAM pH sensor offering resilience to photobleaching at normal measurement powers (10 µW, black points), and some degradation at higher power (100 µW, red points). **(B)** "Standard" loaded FAM sensor showing much greater photobleaching under same conditions. Representative response to pH is shown by the blue lines for relative scale.

## Repeatability of optrode coupling

Repeatability of sensors loading was investigated through a survey of spectra from fibre cores. Figure S3 A shows sensor signal from 4 different cores from multiple fibres for the pH sensor and oxygen sensor (S3 B). A repeatable sensor signal was obtained from different cores across multiple fibres for both the pH and oxygen sensor with some variation in amplitude likely due to varying sized microspheres with differing seating and therefore optical coupling in the cores. Distinct spectral characteristics allows identification of differing sensors locations.

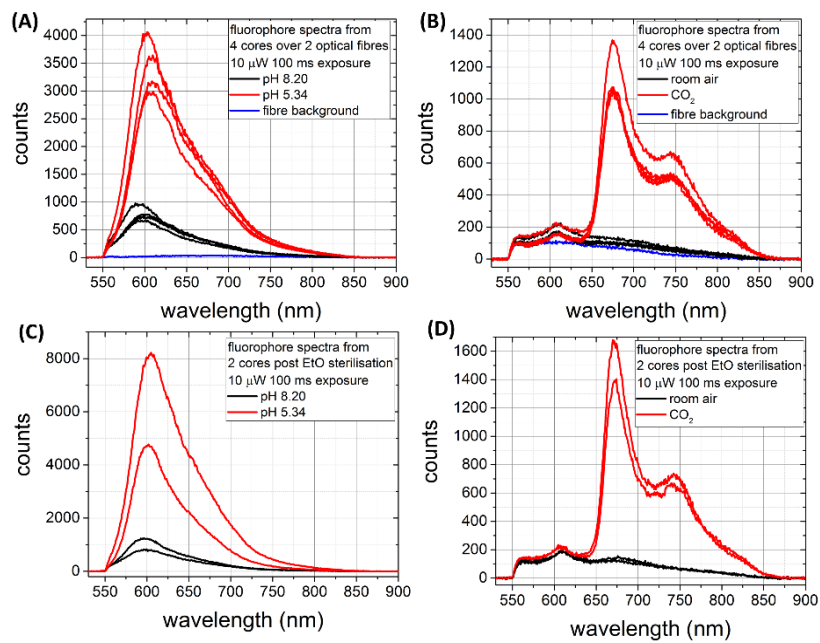

Figure S3. Spectra of pH and O2 sensors with varying conditions from multiple optrode. All the measurements were taken at 520 nm laser (10 µW, 100 ms). (A) shows signal from four different cores from two optrode with pH sensor. Red line represents spectra at pH 5.3 and black line is the spectra obtained at pH 8.2. The blue line is the fibre background in absence of any sensor. (B) shows pH response of two cores at different pH after EtO sterilisation. (C) shows signal from four different cores from two fibres with oxygen sensor loaded. Red line represents spectra at no oxygen (depleted by $CO_2$) and black line is the spectra obtained at room air (21% oxygen). The blue line is the fibre background in absence of any sensor. (D) shows oxygen response of two cores after EtO sterilisation.

## Stability after sterilization

For clinical application, any sensing optrode must undergo and survive suitable sterilisation process. The optrode with pH and oxygen sensor were prepared and sterilised using ethylene oxide (EtO) sterilisation process, which is commonly used for medical devices sterilisation. The optrode performance was tested before and after sterilisation (Figure S3). A survey of cores was performed and spectra was recorded and compared pre and post sterilisation. The sensors survived suitable sterilisation regime used for clinical devices but further tests are required such as toxicological and biocompatibility studies.